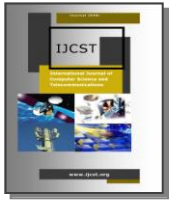

# Agent Approach in Support of Enterprise Application Integration

Djamel Benmerzoug
Faculty of NTIC, University of Constantine 2
benmerzougdj@yahoo.fr

*Abstract*— The present approach highlights the synergies between application integration and interaction protocols. Since both fields have advanced in different directions, a number of important technical problems can be addressed by their proper synthesis.

In our previous work [7][6][5], we proposed a methodological approach based on Interaction Protocols for Enterprise Application Integration (EAI). This approach permits to specify MAS (Multi-Agent System) interaction protocols, verify their behavior and use them to integrate multiple business applications.

The result of the proposed approach is a validated interaction protocol. Based on this protocol, we define in this paper, an agent-based architecture for the EAI. It includes all the concepts necessary to support communication and coordination mechanisms such as inter-agent and agent-Web services communication.

## I. INTRODUCTION

THE penetration of Internet and the World Wide Web in accordance with new technological advances urged companies to seize the opportunities offered by electronic commerce and electronic business. Especial promising are the opportunities for cooperation among several geographically distributed orga- nizations based on open networks and innovative information and communication technologies. Indeed, several companies have showed many interests to the integration of these tech- niques.

Interaction Protocols (IP), provide a middle ground between rigid automation and flexible manual execution. Using protocols as a mechanism to structure communications, agents can be less sophisticated. Protocol designers analyze and design protocols for desirable properties. Agents publicly declare the protocols in which they can participate making it easier to find agents with whom to interoperate.

The importance of interaction is increasing as more and more monolithic systems are broken down into smaller services. The importance of interaction has been stressed by many authors [23][7][12]. Moreover, interaction is also considered a key issue in various industry standard[22].

In previous work [6][7], we described the use of IP to define and manage collaborative processes in B2B (Business to Business) relationships where the autonomy of participants is preserved. We showed the practicability of our approach by embedding it in a Web services language for specifying protocols, which conducive to reuse, refinement and aggregation of our business protocols.

We also elaborated translation rules from interaction protocols notations used in our approach into Colored Petri Nets (CPN). These rules are implemented in IP2CPN[1] [7]: the tool we developed to automatically generate Petri nets from protocols specifications. Resulting Petri nets can be analyzed with dedicated tools to detect errors as early as possible.

In this paper, we describe how the Multi-agent System (MAS) use the verified and the validated IP to establish the EAI. The proposed architecture based on the notion of agent. It includes all the concepts necessary to ensure all the phases of the EAI life-cycle. Our suggestion consists in the addition of a specific agent between the MAS application and its IP parts conceived as Web services. In fact, we defined two types of agent, namely, the Enterprise Agent representing an individual enterprise and the Integrator Agent.

The remainder of the paper is organized as follows: Section 2 discusses some related work. Section 3 overviews our previous work : The use of IP to define and manage collaborative processes in B2B relationships. Section 4 and 5 give details about the proposed agent-based architecture. In section 6, we study the different aspects related to the communication in our architecture. Section 7 gives some implementation aspects. Finally, concluding remarks are given in section 8.

## II. RELATED WORK

EAI is an active research domain. The community is still debating the issues of enterprises collaboration at the business process level. In fact, several areas of research are relevant to our work. We discuss some of them briefly and highlight the differences.

a) *Web services composition*.: BPEL4WS [17] is a language designed to specify the static composition of Web services. However, it mixes interaction activities and business logic making it unsuitable for reuse [13]. In contrast to our approach, BPEL4WS elements are only used to specify messages exchanges between the different business partners. Afterwards, this specification is used by agents to enact the integration of business processes at run time.

The Symphony project [10] has developed an algorithm for analyzing a composite service specification for data and control dependences and partitioning it into a set of smaller components. These components are then distributed to different locations and, when deployed, cooperatively deliver the same semantics as the original workflow. Symphony does not provide any support for failures arising from workflow mismatches since it assumes that the distributed processes will be derived from a single complete BPEL process.

---

[1] IP2CPN: Interaction Protocols To Colored Petri Nets

Other approaches try to combine Web services with multi-agent systems to ensure composition and interoperability. For example [8] summarize the relationship between agents and Web services with the aphorism Adaptive Workflow Engines = Web Services + Agents: namely, Web services provide the computational resources and agents provide the coordination framework. They propose the use of the BPEL4WS language as a specification language for expressing the initial social order of the multi-agent system. [8] does not provide any design issues to ensure the correctness of their interaction protocols.

b) *Interaction protocols modelling.*: Interaction protocols are specific, often standard, constraints on the behaviors of the autonomous agents in a multiagent system. Protocols are essential to the functioning of open systems, such as those that arise in most interesting Web applications.

The Dooley graph [21] is an alternative formalism for visualizing agent inter-relationships within a conversation. Object-oriented methods like UML [15] offer a way to reduce the gap between users and analyst when considering message transfers, yet they only address the dynamic behavior of individual objects and are informal.

The state transition diagram has been extensively used for IP specification due to its clarity. The weakness is that it does not reflect the asynchronous character of the underlying communication [19]. Furthermore, it is not easy to represent integration of protocols.

To ensure the correctness of IP specification at design time, [16] propose a translation rules for the conversation of an interaction protocol given in AUML to Petri net. Unfortunately, no automatic procedures were provided that guide the conversation of an interaction protocol given in AUML to Petri net representations.

c) *Discussion.*: As shown before, several important strands of research of relevance to interaction protocols have been carried out before. However, the synthesis, enhancement, and application of these techniques on business process integration is a novel contribution of this paper.

Given the autonomy and heterogeneity of business partners, we believe that the most natural way to treat interactions is as communications. An interaction protocol involves the exchange of messages with a streamlined set of tokens. Indeed, the use of IP to define business process integration enables a greater autonomy of companies because each company hides its internal activities, services and decisions required to support public processes.

In fact, our approach allows us to provide a clear separation of the inter-enterprise collaboration management and local business process one. This dichotomy makes full use of existing workflow system components, to support both public processes and private business processes.

Formalisation and verification of autonomous and heterogeneous collaborative business processes is a key prerequisite for the effective business application. As asserted by [20], minimal specifications based on external behavior will maximize interoperability. The present approach shows a design issues of how IP can be specified and verified.

Another advantage of our approach is the integration completeness property inherent from our BPEL4WS specification. It means that the IP is itself published and accessed as a Web service that can participate in other application integration. Particular protocols may then be selected and composed to support a desired business application.

### III. AN OVERVIEW OF THE PROPOSED APPROACH

In recent years, EAI modelling and reengineering have been longstanding activities in many companies. Most internal processes have been streamlined and optimized, whereas the external processes have only recently become the focus of business analysts and IT middleware providers. The static integration of inter-enterprise processes as common in past years can no longer meet the new requirements of customer orientation, flexibility and dynamics of cooperation [2].

In [4] we have developed an agent-based method for developing cooperative enterprises information systems. This method permits to explicitly map the business process into software agents. In [5], we have described the use of IP to define and manage public processes in B2B relationships.

The use of IP to define public processes enables a greater autonomy of companies. In this way, the IP provide a high abstraction level in the modelling of public processes.

In other hand, the B2B integration scenarios typically involve distributed business processes that are autonomous to some degree, hence the importance of IP-based modelling. IP are a useful way for structuring communicative interaction among business partners, by organizing messages into relevant contexts and providing a common guide to the all parts. Formally an IP is defined as follow:

Definition : An Interaction Protocol is a quadruplet:

$$IP = <ID, R, M, f_M>, \text{ where:}$$

- ID is the identifier of the interaction protocol
- $R = r_1, r_2, \ldots, r_n$ ($n > 1$) is a set of Roles (private business process or Web Services)
- M is a set of non-empty primitive (or/and) complex messages, where:
  - A Primitive Message (P M) corresponds to the simple message, it is defined as follow:
    P M = <Sender, Receiver, CA, Option>, where:
    * Sender, Receiver $\in$ R
    * CA $\in$ FIPA ACL Communicative Act (such as: cfp, inform, ...)
    * Option: contain additional information (Syn- chronous / Asynchronous message, constraints on message,...)
  - A Complex Message (CM) is built from simpler (primitive) ones by means of operators:
    $CM = P M_1 \text{ op} P M_2 \ldots \text{op} P M_m$, where:
    * $m > 1$, op $\in$ {XOR, OR, AND}, and

* $\forall\ i \in [1,m[,\ PM_i.Sender = PM_{i+1}.Sender,$ $PM_i.Sender \in R.$

- $f_M$ : a flow relation defined as : $f_M \subseteq$ (RxR), where (RxR) is a Cartesian product $(r_1, r_2) \in$ (RxR), for $r_1, r_2 \in R$

Figure 1 shows our approach for the treatment of business process integration based on interaction protocols. In this approach, we find two levels:

- Design level: Formalisation and verification of interaction protocols.
- Operational level: Management of interaction protocols at run-time.

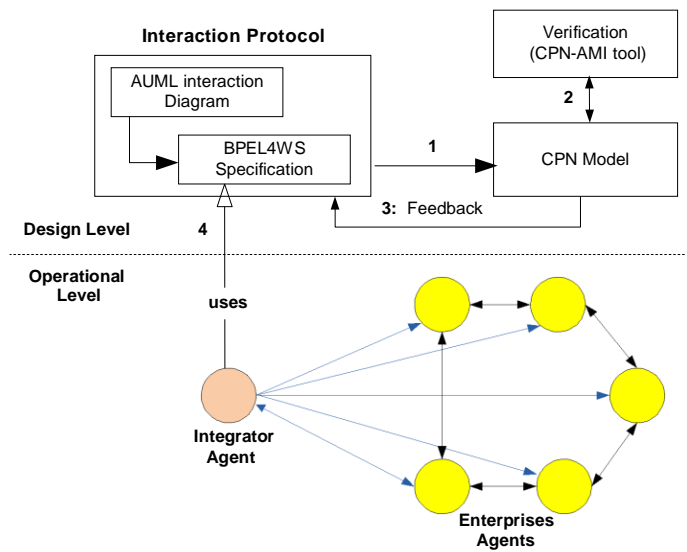

Fig. 1. The proposed approach

### A. Design level: Formalisation and Verification of Interaction Protocols

Developing effective protocols to be executed by autonomous partners is challenging. Similar to protocols in traditional systems, IP in open and web-based settings need to be specified rigorously so that business partners can interact successfully.

For this reason, developed a method for IP design and verification. This method (see figure 1) uses different models and languages.

Our method motivates the use of IP based on AUML/BPEL4WS for application integration modelling, where pre- and post-conditions, rules, guards are specified in OCL[2].

AUML (Agent UML) notation [1] [3] is a UML profile dedicated to agents trying to simplify the transition from software engineering to multi-agent system engineering. In other hand, BPEL4WS [17] (Business Process Execution Language for Web Services) is a de facto standard for describing Web services composition. In our context, BPEL4WS was used as a specification language for expressing the interaction protocols of the multi-agent system [5].

When protocols are employed in open environments, such as the internet, they must be executed by agents that behave more or less autonomously and whose internal designs are not known. In such settings, therefore, there is a risk that the participating agents may fail to comply with the protocol [18]. Without a rigorous means to verify compliance, the very idea of protocols for interoperation is subverted.

The use of formal methods is important because ensuring the correctness of complex protocols is seldom possible via other design approaches. High-level Petri nets are a suitable formal method for the design of IP because of their ability to express concurrency, non-determinism and system concepts at different levels of abstraction. CPN are a popular form of high-level Petri nets that have extensive tool support for the design of systems, including protocols.

For this reason, we use Colored Petri Net (CPN) as target notation. CPN allows analysis to check for properties in IP. The semantics of the IP notations used in our approach and its application are described on the basis of translation rules into CPN. Consequently, lifelines, messages, constraints, splitting/merging paths, interaction terminations, and other IP construction elements are translated into a Petri net. Afterwards, the resulting Petri net specification can be analysed by dedicated tools to detect errors as early as possible.

In a previous work [6], we elaborated translation rules from interaction protocols notations used in our approach into Colored Petri nets. These rules are implemented in IP2CPN: the tool we developed to automatically generate the Petri net from protocols specification. The resulting Petri net specification can be analyzed by CPN-AMI tool [11] to detect errors as early as possible.

### B. Operational Level: Management of interaction protocols at run-time

As we already have said before, the BPEL4WS process specification is considered as a specifying language for expressing the interaction protocol of the multi-agents system. In this the rest of this paper we briefly describe how the MAS use the verified and the validated BPEL4WS specification to establish the EAI. Our suggestion consists in the addition of a specific agent between the MAS application and its IP parts conceived as Web services.

### IV. AN AGENT-BASED ARCHITECTURE FOR ENTERPRISE APPLICATION INTEGRATION

The proposed architecture based on the notion of agent. It includes all the concepts necessary to ensure all the phases of the Enterprise Application Integration life-cycle. As shown in Fig. 2, our suggestion consists in the addition of a specific agent between the MAS application and its IP parts conceived as Web services. In fact, we defined two types of agent, namely, the Enterprise Agent representing an individual enterprise and the Integrator Agent.

---

[2]OCL: Object Constraint Language. (www.omg.org/cgi-bin)

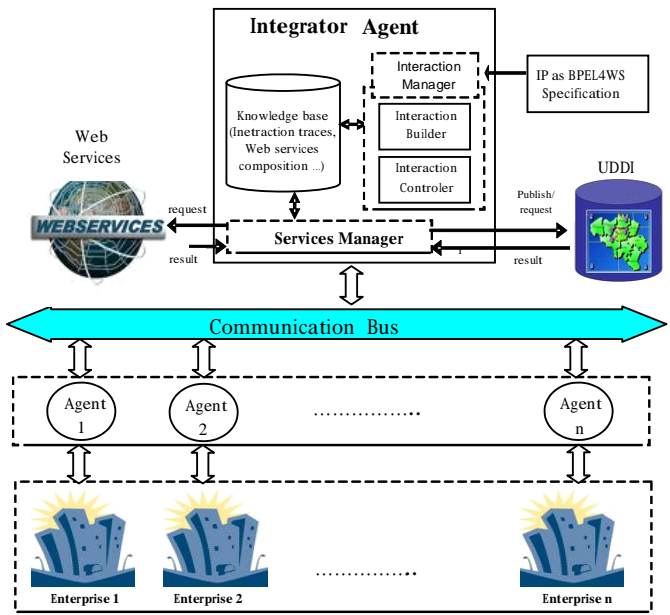

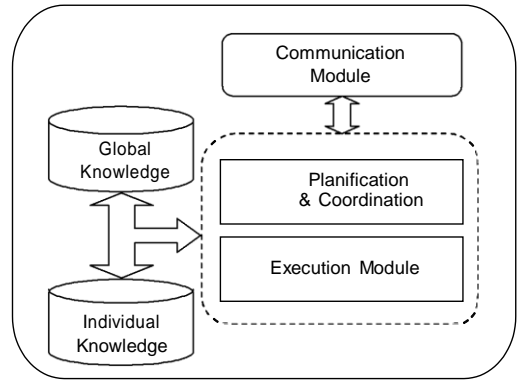

Fig. 2. The proposed architecture

Fig. 3. The structure of the Enterprise Agent

The main roles of the Integrator agent are the creation, monitoring, and control of IP life cycle. Its architecture features two modules: an interaction manager and a service one. The interaction manager contains operational knowledge (e.g., Interactions states). It also provides operations for monitoring interaction (i.e., creating and deleting instances). The service manager provides methods for receiving service requests, tracing service executions, and communicating with service requesters in accordance with IP definition (e.g., sending a notification informing the requester that deadline for cancelling an operation is passed).

A. Description of the Integrator Agent

As shown in Fig. 2, the Integrator agent is composed of four main modules.

- The knowledge module includes a database and a set of actions. The role of this module is to store information that is necessary for the success of the interaction. Such information is traces of interaction used during the phases of the current states and coordination tasks running. This information is useful for future use of the protocol.
- The Interaction Manager module allows the processing of messages depending on the specification of the interaction protocols (the BPEL4WS specification) and the knowledge parts.
- The services management module provides the services necessary for locating and invoking services.
- The communication module allows the transport of messages. In our work, we use the FIPA-ACL [14] as the language of communication between the different agents. This module has the role to structure the messages constructed by the agent.

B. Description of the Enterprise Agent

We consider the Enterprise Agent as an entity including a communication module, a planning and coordination module, and an execution module (Fig. 3).

- The Communication Module: contains all the processes required to handle the messages: reception, filtering, and translation of incoming messages, and formulation and sending of the outgoing messages.
- The Planning and Co-ordination Module: is responsible for managing the cooperation and formulating the offers for achieving sub-goals announced by the Integrator agent.
- The Execution Module: this module contains the information about the internal resources of the individual enterprise, which makes possible the performance of local tasks that are assigned to the enterprise. It establishes the correspondence between the sub-goal assigned to the agent and the internal resources of the enterprise capable of achieving this sub-goal.
- The Global Knowledge Module: contains information related to the organisational and operational rules defined by the enterprises.
- The Individual Knowledge Module contains information about the agent itself: its capacities and skills, and the current state and workload, i.e. for each skill, indicators are assigned to determine availability, as well as the cost of such skill.

V. ROLES AND BEHAVIORS OF AGENTS

IP must specify the semantics of the messages exchanged. It also describes the reactions of an agent in a given message. In our case, two main roles involved in the interaction protocols:

1) The Integrator agent supports data extraction from the IP specification and message exchange between agents involved in the interaction.

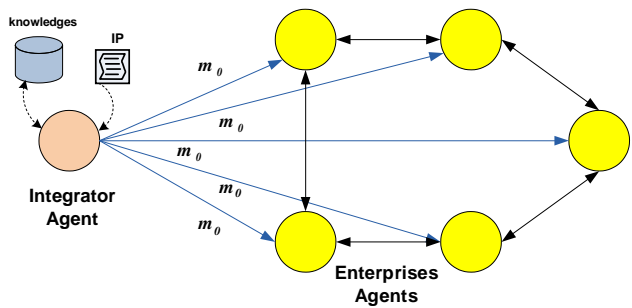

Fig. 4. Agents behaviors

2) The role of participants' agents is to coordinate to satisfy all the requests of the Integrator agent. The interaction protocol that we specified in this paper describes the rules used by the participants' agents for a correct coordinate.

The behavior of agents, as shown in Figure 4, can then be summarized as follows:

1) The Integrator agent (the initiator of the IP) announces the beginning of the integration process and the allocation of tasks. For each task to be performed, the Integrator agent announces a specification that includes a description, and a list of requirements and constraints for this task. This information is encapsulated in a message, which represents the initiator message ($m_0$ in Fig. 4).
2) When participants' agents are identified, the allocation phase of sub goals for participating agents begins. Each participant (Agent) performs the part that has been assigned. The composition of individual results leads to the realization of the overall goal.
3) The Integrator agent ensures the coordination between the different enterprises agents in accordance with the rules of interaction defined in the BPEL4WS specification. In the context of this work we will not discuss the aspect of local coordination, which may be different from one company to another.

To ensure the successful completion of the interaction, the Integrator agent must know how many responses should be expect from the participating agents. The analysis of interaction rules defined in the BPEL4WS specification and the semantics of the ACL allow when if others messages may be received or not.

## VI. COMMUNICATION MODES

This integration process must keep as much as possible the autonomy of architecture core based on agents. Indeed, The agents are coordinated with the Integrator agent and the exchange of messages to enact the EAI. In this architecture, the following communications pathways exist:

- agent to agent communication occurs via FIPAs Agent Communication Language (ACL) and is facilitate by a FIPA compliant Agent Management System.

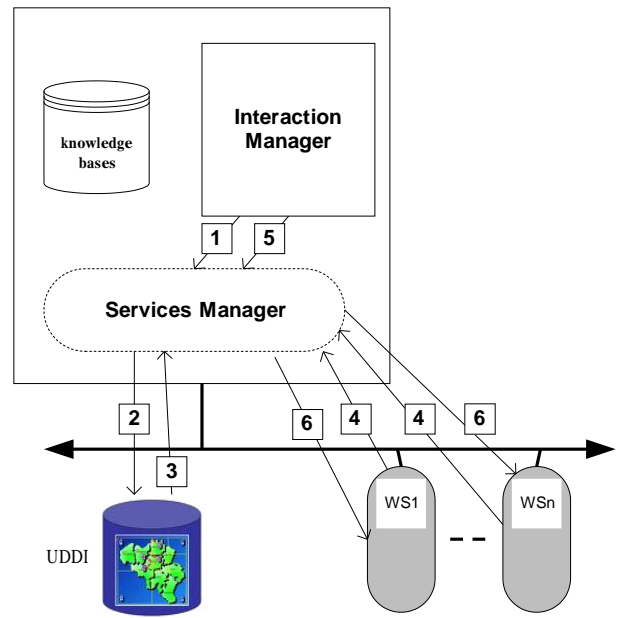

Fig. 5. Agent-Web Services Communication

- agent to Web service communication is accomplished via SOAP messages.
- agent to BPEL4WS dataspace communication utilizes appropriate protocols/interfaces provided by the dataspace. The dataspace is used to store BPEL4WS process variables, which maintain the state of the IP.

### A. Inter-Agent Communication

FIPA [14] supplies a standard syntax for messages. These messages are based on the theory of the act of speech, which is the result of the linguistic analysis of human communication. The basis of this theory is to produce an action from the language. In the FIPA-ACL, no specific language for the description of the contents of messages is imposed. Several languages can be used for the description of the contents of the exchanged messages such as KIF (Knowledge Interchange Format), Semantic Language (SL), Prologue and XML (eXtensible Mark-up Language). XML will be used for the description, the specification and the interpretation of the contents of messages exchanged. So, the messages exchanged among the agents are described in FIPA-ACL/XML. The use of XML for the contents of communications among agents permits the display of messages in a Web browser and facilitates the integration with existing Web-based applications.

### B. Agent - Web service Communication

This type of communication is ensured by the Services Manager module of the Integrator agent. This module provides the necessary services for locating and invoking services. Its behavior is presented as follows (see Fig. 5):

1) When the Integrator agent detects a need for which it does not have the capacity, it uses the functionality of

the Service Manager module to find Web services that could satisfy his need.
2) The Service Manager module requests the UDDI registry
3) The Service Manager module obtains from the registry a list of descriptions of Web services that matching the search criteria.
4) The Service Manager module starts the invocation of services found to obtain their attributes.
5) The Integrator agent requests the service manager module to invoke the Web service selected.
6) The Service Manager module sends a parallel invocation message to the selected service and cancel messages to other services not concerned.

## VII. SOME IMPLEMENTATION ASPECTS

The platform chosen for the implementation is Java Agent Development framework (JADE) [9]. JADE is a software development framework, fully implemented in Java, which aims at the development of multi-agent systems and applications that comply with FIPA standard. To achieve such a goal, JADE offers the following list of features to the agent programmer:

- Distributed agent platform. The agent platform can be distributed on several hosts, each one of them executes one Java Virtual Machine.
- FIPA-Compliant agent platform, which includes the Agent Management System, the Directory Facilitator and the Agent Communication Channel.
- Efficient transport of ACL messages between agents.

### A. Using JADE to Develop the Proposed Architecture

Creating a JADE agent is as simple as defining a class extending the Jade.core.Agent class and implementing the setup() method. The setup() method is intended to include agent initializations. The actual job an agent has to do is presented as JADE behaviors. The Fig. 6 represents the implementation of the Integrator and the Enterprise agent classes. These classes are extensions of the basic Agent class defined in JADE.

The communication module presented in our architecture is derived from JADE class CyclicBehaviour and will therefore run continuously. If no messages have arrived, the behavior will block and restart after a new message has arrived. If a message has arrived, the perception interface has to interpret this message into a set of goals, starts the execution module and resumes waiting for incoming messages. The communication module behavior code is shown in Fig. 6.

The Interaction Manager Module is derived from the class SimpleBehaviour. This module takes as input a BPEL4WS specification and builds the appropriate interaction protocol. In this context, an interaction protocol is implemented as an FSMBehaviour behavior. The FSMBehaviour class is a CompositeBehaviour that executes its children (sub behaviour that are going to be used by FSMBehaviours) according to a Finite State Machine defined by the user. In fact, each sub-behaviour is dedicated to an interaction that represents the set of messages exchanged to satisfy a given set of requirements.

```
public class Integrator extends Agent {
    protected void setup() {
addBehaviour(new SimpleBehaviour(this)
{
            // Traitement ...
        }
}
}

public class EntrepriseAgent extends
Agent {
    class recevoir extends
SimpleBehaviour {
        //Traitement ...
    }
    public recevoir(Agent a)
    {
        super(a);
    }
    protected void setup() {
        recevoir mybehaviour = new
recevoir(this);
        addBehaviour(mybehaviour);
    }}
         - a -
```

```
import java.util.*;
import jade.core.*;
import jade.core.behaviour.*;
import jade.lang.acl.ACLMessage.*;
class Communication extends CyclicBehaviour
{
    public Communication (Agent a)
{super(a);}
    Vector goals = new Vector();
    public void action() {
    // wait for message
    ACLMessage received =
        myAgent.receive();
    if(ACLMessage == null) {block();}
    else {
    // message interpretation
    ...................
    // Start PlanRetrieval Behaviour
    addBehaviour(new
    PlanRetrieval(this,goals));
} } }
         - b -
```

Fig. 6. a) Agent Class Extension, b) Partial specification of the Communication module

## VIII. CONCLUSION

In this paper, we have proposed an agent-based architecture for enterprise application integration. The multi-agent provides high level communication languages that preserve the entire richness of the communications. Thus, the FIPA-ACL is used in our approach to formulate messages exchanged between the different agents constituting the system. The content of these messages is described with XML. The use of the JADE platform has allowed us to implement the Integrator Agent and the Enterprise Agents and thus, to simulate the interaction between these entities.